\documentstyle[sprocl]{article}

\input{psfig}

\def\Journal#1#2#3#4{{#1} {\bf #2}, #3 (#4)}

\def\NPA{{\em Nucl. Phys.} A}
\def\NPB{{\em Nucl. Phys.} B}
\def\PLB{{\em Phys. Lett.}  B}
\def\PRL{\em Phys. Rev. Lett.}
\def\PRC{{\em Phys. Rev.} C}
\def\PRD{{\em Phys. Rev.} D}
\def\ZPA{{\em Z. Phys.} A}

\def\be{\begin{equation}}
\def\ee{\end{equation}}
\def\bea{\begin{eqnarray}}
\def\eea{\end{eqnarray}}
\begin{document}

\title{OBSERVABLES IN UNPOLARIZED AND POLARIZED VIRTUAL COMPTON
SCATTERING}

\author{A. METZ}

\address{Institut f\"ur Theoretische Physik, Ruprecht Karls--Universit\"at
\\ Philosophenweg 19, D--69120 Heidelberg, Germany
\\ E-mail: metz@frodo.tphys.uni-heidelberg.de}

\author{B. PASQUINI}

\address{Institut f\"ur Kernphysik, Johannes Gutenberg--Universit\"at
\\ J. J. Becher-Weg 45, D--55099 Mainz, Germany
\\ E-mail: pasquini@kph.uni-mainz.de}

\author{D. DRECHSEL}

\address{Institut f\"ur Kernphysik, Johannes Gutenberg--Universit\"at
\\ J. J. Becher-Weg 45, D--55099 Mainz, Germany
\\ E-mail: drechsel@kph.uni-mainz.de}


\maketitle\abstracts{Below pion threshold virtual Compton scattering
off the nucleon gives access to the generalized electromagnetic
polarizabilities.
Different theoretical results for the generalized polarizabilities
have been compared.
In particular, the influence of the generalized polarizabilities
on the unpolarized cross section and on double polarization 
observables has been investigated.
Predictions for these observables have been obtained in
the linear sigma model and chiral perturbation theory.}

\section{Introduction}
Compton scattering has a long history 
as an interesting tool to investigate
the nucleon's structure.
Though it has been applied in different kinematical 
regions, Compton scattering is of particular interest below the threshold of 
pion production, where the electromagnetic polarizabilities of the 
nucleon can be studied.
While in the past the main focus was on real Compton scattering (RCS),
recently much effort has been devoted to virtual Compton scattering 
(VCS), $\gamma^{\star} + N \to \gamma + N$, with a virtual photon
$\gamma^{\ast}$ in the initial state and a real photon $\gamma$ in the final 
state.
In comparison to RCS, the VCS reaction contains much more
information because of the variable mass and the additional
longitudinal polarization of the virtual photon.
\\
\begin{figure}
\label{fig_born}
\centerline{
\psfig{figure=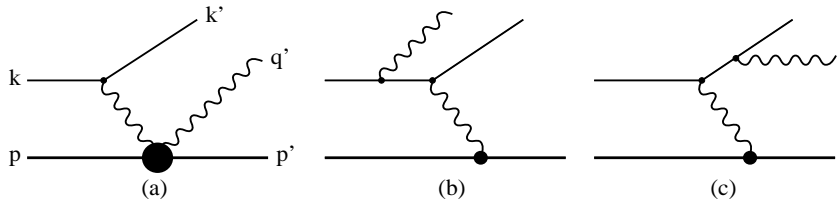,width=11cm}}
\caption
{Feynman diagrams of the reaction
$e(k) + N(p) \to e'(k') + N'(p') + \gamma(q').$ 
(a): the VCS contribution containing the Born and residual 
amplitudes. 
(b) and (c): Bethe-Heitler scattering.}
\end{figure}
As shown in Fig.~\ref{fig_born}, VCS can be realized experimentally by 
electron-nucleon bremsstrahlung, $e + N \to e' + N' + \gamma$.
Besides the VCS mechanism, the bremsstrahlung contains the 
Bethe-Heitler scattering (BH), where the real photon is emitted from
the incoming or outgoing electron.
Therefore, the total amplitude $T^{ee'\gamma}$ is
the coherent sum of the VCS amplitude $T^{VCS}$ and the BH amplitude
$T^{BH}$,
\begin{equation} \label{interf}
T^{ee'\gamma} = T^{VCS} + T^{BH} \,.
\end{equation}
If not stated differently, we use the {\it cm} frame of the final
state photon and nucleon throughout this article.
The 4-momenta of the incoming photon and nucleon are denoted by 
$q^{\mu} = (\omega, \vec{q}\,)$ and $p^{\mu} = (E, -\vec{q}\,)$, respectively, 
while the outgoing particles are characterized by
$q^{\prime \mu} = (\omega', \vec{q}\,')$ and 
$p^{\prime \mu} = (E', -\vec{q}\,')$.
We also define $Q^2 \equiv - q^2$ and $\bar{q} \equiv |\vec{q}\,|$.
For the following discussion it is convenient to introduce the quantity 
\begin{equation} \label{var}
Q_{0}^{2} \equiv Q^{2}|_{\omega'=0}
 = 2m_{N} \Bigl[ \sqrt{\bar{q}^{2} + m_{N}^{2}} - m_{N} \Bigr] \,.
\end{equation}
According to Eq.~(\ref{var}), the variable $Q_0^2$ can be replaced by
$\bar{q}$ and vice versa.

Below pion threshold, the analysis of both RCS and VCS is based on
low energy theorems, which rely on fundamental symmetries as
Lorentz invariance, gauge invariance and crossing symmetry.
The decomposition of the scattering amplitude into a Born 
contribution and the residual part plays an important role in the
derivation of these theorems.
In the case of VCS we~have
\begin{equation} \label{decomp}
T^{VCS} = T_{B}^{VCS} + T_{R}^{VCS},
\end{equation}
where the Born amplitude
$T_{B}^{VCS}$ shows the typical $1/\omega'$ singularity of 
bremsstrahlung, while the residual amplitude $T_{R}^{VCS}$ is 
proportional to $\omega'$ in the low energy 
limit~\cite{Guichon_95,Scherer_96}.
The Born amplitude is entirely determined by the nucleon mass, its
charge and the electromagnetic form factors, $G_E$ and $G_M$, i.e.,
by definition $T_{B}^{VCS}$ contains  only properties of the nucleon
in its ground state.
The influence of the excitation spectrum enters through
the residual amplitude which can be parametrized by the 
generalized polarizabilities (GPs) of the nucleon~\cite{Guichon_95}.
To leading order in $\omega',$ there appear 10 GPs in $T_{R}^{VCS}$.
These GPs depend on $Q_0^2$ or $\bar{q}$ and are related with 
the electromagnetic polarizabilities of RCS
whose leading contribution is completely determined by the well known
electric ($\alpha$) and magnetic ($\beta$) polarizabilities.
We discuss the GPs in the following section in more details.

The measurement of the GPs requires data with a very high accuracy 
which can only be obtained by means of the new electron 
accelerator facilities like MAMI, MIT-Bates and Jefferson Lab.
The VCS reaction has already been measured at  MAMI~\cite{MAMI_prop}
for $\bar{q} = 600$ MeV, it will be investigated at $\bar{q} = 240$ MeV at 
MIT-Bates~\cite{Bates_prop}, while the activities at 
Jefferson Lab~\cite{TJNAF_prop} will concentrate on the region of 
$\bar{q} \approx 1$ GeV.
Very recently, the first preliminary results for $\alpha(\bar{q})$ 
and $\beta(\bar{q})$ have been extracted from the 
MAMI experiment~\cite{Walcher_98}.
\\
On the theoretical side the GPs were predicted by various 
approaches.
Calculations of all GPs have been performed in the nonrelativistic 
constituent quark model (CQM)~\cite{Guichon_95,Liu_96}, chiral 
perturbation theory in the heavy baryon formalism (HBChPT) including 
the $\Delta$ resonance as a dynamical degree of 
freedom~\cite{Hemmert_97a,Hemmert_97b,Knoechlein_97}, and in the 
one-loop approximation of the linear sigma model 
(LSM)~\cite{Metz_96,Metz_97}.  
The $\bar{q}$ behaviour of $\alpha$ and $\beta$ was determined in an 
effective Lagrangian model~\cite{Vanderhaeghen_96} containing several
resonances and the exchange of $\pi^0$ and $\sigma$ mesons in the 
$t$ channel.
In addition, there exists a Skyrme model prediction of 
$\alpha(\bar{q})$~\cite{Kim_97}, while the paramagnetic part 
of $\beta(\bar{q})$ was
calculated in a relativistic quark-model formulated in the light-front 
dynamics~\cite{Pasquini_98}.
\\
The influence of the GPs on observables as the unpolarized
cross section of the reaction $p (e, e' p') \gamma$ and asymmetries
for beam-recoil polarization, $p (\vec{e}, e' \vec{p}\,') \gamma,$ has 
been studied in the effective Lagrangian approach by 
Vanderhaeghen~\cite{Vanderhaeghen_96,Vanderhaeghen_97}.
In these studies  the contributions to the amplitude $T_{R}^{VCS}$ 
were investigated to all orders in $\omega',$ i.e., the effects
of higher order GPs were automatically incorporated.
In contrast to this, we will only consider the dipole approximation to
$T_{R}^{VCS}.$ 
Regarding the validity of this approximation, see the discussion in sec.~2.1.
In the unpolarized case, the $\omega'$ dependence of the effect due to 
the GPs was previously calculated in HBChPT~\cite{Knoechlein_97}. 
In addition to unpolarized cross sections, we also predict asymmetries
of beam-target polarization, $\vec{p} (\vec{e}, e' p'\,) \gamma,$
which can be explored at Bates using the BLAST target-detector system.
We will show predictions of the LSM and HBChPT focusing on the influence
of the GPs as function of the $\gamma^{\ast}\gamma$ scattering angle.
\section{Generalized Polarizabilities}
\subsection{Definitions and Constraints}
The concept of GPs was first introduced in connection with nuclear
targets~\cite{Arenhoevel_74} and has recently been discussed in detail 
for the specific case of the nucleon~\cite{Guichon_95}.
The GPs can be defined through the multipoles
$H_{R}^{(\rho' L',\rho L)S}(\omega',\bar{q})$ 
of the residual amplitude.
In the notation of the multipoles, $\rho \,(\rho')$ denotes the
type of the incoming (outgoing) photon 
($\rho=0$: Coulomb, $\rho=1$: magnetic, $\rho=2$: electric)
while $L \,(L')$ refers to the angular momentum of the initial (final) photon. 
The quantum number $S$ indicates a no spin--flip ($S=0$) or a 
spin--flip ($S=1$) transition.
\\
In order to define the GPs it is suitable to replace
all electric transitions. As a consequence one introduces the socalled mixed 
multipoles $\hat{H}_{R}^{(\rho' L',L)S}$  describing 
a well defined mixture of an electric and a charge transition
in the initial state.   
More details on this technique can be found in the 
literature~\cite{Guichon_95,Guichon_98}.
The GPs are finally given by
\begin{eqnarray}
P^{(\rho' L',\rho L)S}(\bar{q}) & = &
 \biggl[ \frac{1}{\omega^{\prime L'} \bar{q}^{L}}
   H_{R}^{(\rho' L',\rho L)S}(\omega',\bar{q}) \biggr]_{\omega'=0}
 \quad (\rho,\rho' = 0,1) \,, 
 \nonumber \\
\hat{P}^{(\rho' L',L)S}(\bar{q}) & = &
 \biggl[ \frac{1}{\omega^{\prime L'} \bar{q}^{L+1}}
   \hat{H}_{R}^{(\rho' L',L)S}(\omega',\bar{q}) \biggr]_{\omega'=0}
 \quad (\rho' = 0,1) \,,
\end{eqnarray}
and carry the same quantum numbers as the corresponding 
multipoles~\cite{Guichon_95}.
\\
In the dipole approximation for the real photon 
($L'=1$), selection rules due to parity and angular momentum 
conservation lead to the ten possible transitions listed in 
Table~\ref{tab_pol},
3 scalar GPs ($S=0$) and 7 vector 
GPs ($S=1$).
\begin{table}[t]
\label{tab_pol}
\caption{Possible transitions to VCS in the case of the dipole 
 approximation for the real photon (E: electric, M: magnetic, C: Coulomb).
 The last column gives the notation of the GPs.} 
\vspace{0.3cm}
\begin{center}
\begin{tabular}{|c|c|c|c|} \hline
 final photon & initial photon & $S$ & GPs \\ \hline
 E1 & C1 & $\hphantom{A}$ 0,1 $\hphantom{A}$ & $P^{(01,01)S}$ \\ 
 E1 & E1 & 0,1 & $P^{(01,01)S},\;\hat{P}^{(01,1)S}$ \\ 
 E1 & M2 & 1 & $P^{(01,12)1}$ \\ \hline
 M1 & M1 & 0,1 & $P^{(11,11)S}$ \\ 
 M1 & C0 & 1 & $P^{(11,00)1}$ \\ 
 M1 & C2 & 1 & $P^{(11,02)1}$ \\ 
 M1 & E2 & 1 & $P^{(11,02)1},\;\hat{P}^{(11,2)1}$ \\ \hline
\end{tabular}
\end{center}
\end{table}
Note that the dipole approximation is equivalent to keeping only
the leading (linear) order in $\omega'$ for the residual
amplitude.
Since the relevant expansion parameter is $\omega'/m_{\pi},$ 
the dipole approximation describes the experimental data 
for RCS only if 
$\omega' \le 60$ MeV, well below  the pion mass $m_\pi.$
However, in VCS the situation is more complicated.
As has been pointed out recently, higher order terms in $\omega'$
become more important if $\bar{q}$ is of the same order of 
magnitude as $\omega'$~\cite{Drechsel_98b}.
Therefore, beyond the restriction given by the pion mass, the condition
$\bar{q} \gg \omega'$ should be fulfilled to guarantee the validity of
the dipole approximation in VCS.
\\
In part the GPs are related to the electromagnetic polarizabilities
which govern the low-energy expansion of the RCS amplitude.
The scalar polarizabilities $P^{(01,01)0}(\bar{q})$ and
$P^{(11,11)0}(\bar{q})$ generalize $\alpha$ and $\beta$ to the case
of virtual photons,
\begin{equation} \label{rcs_pola}
P^{(01,01)0}(\bar{q}) = - \frac{4\pi}{e^{2}} \frac{\sqrt{2}}{\sqrt{3}}
 \alpha(\bar{q}) \,, \quad
P^{(11,11)0}(\bar{q}) = - \frac{4\pi}{e^{2}} \frac{\sqrt{8}}{\sqrt{3}}
 \beta(\bar{q}) \;, 
\end{equation}
with $e^2/4\pi\approx 1/137.$
Two of the vector GPs are connected with the four spin polarizabilities 
$\gamma_i$ of RCS as defined by Ragusa~\cite{Ragusa_93}.
The relations read~\cite{Drechsel_98a}
\begin{equation} \label{rcs_polb}
P^{(01,12)1}(0) = - \frac{4\pi}{e^{2}} \frac{\sqrt{2}}{3} \gamma_{3}
 \,, \quad
P^{(11,02)1}(0) = - \frac{4\pi}{e^{2}} \frac{2\sqrt{2}}{3\sqrt{3}}
 (\gamma_{2} + \gamma_{4}) \,. 
\end{equation}
The remaining two combinations of spin polarizabilities of RCS can not be 
related to the GPs in the dipole approximation.
In particular, the forward spin-polarizability 
$\gamma=\gamma_1-\gamma_2-2\gamma_4$ is not contained 
in this kinematical limit of VCS~\cite{Drechsel_98a}.

As has been shown by our explicit calculation in the LSM~\cite{Metz_96}
and later by a model-independent analysis~\cite{Drechsel_97}
based on charge conjugation symmetry and nucleon crossing,
the ten GPs have to satisfy four constraints.
For instance, the three scalar GPs obey the condition 
\begin{equation} \label{pol_zsh}
\frac{e^{2}}{4\pi} \hat{P}^{(01,1)0}(\bar{q}) =
 - \frac{Q_{0}^{2}}{3 m_{N} \bar{q}^{2}}
 \Bigl[ \alpha(\bar{q}) + \beta(\bar{q}) \Bigr] \,.
\end{equation} 
Similarly, there are three relations
among the vector GPs~\cite{Metz_97,Drechsel_98a}.
In addition to these relations, the vector GPs fulfill specific constraints
at $\bar{q} = 0$.
Three vector GPs vanish at the origin, 
\begin{equation} \label{const_a}
P^{(01,01)1}(0) = P^{(11,11)1}(0) = P^{(11,00)1}(0) = 0 \,,
\end{equation}
while the remaining four satisfy the condition
\begin{equation} \label{const_b}
P^{(01,12)1}(0) + \sqrt{3} P^{(11,02)1}(0)
 - \sqrt{3}\hat{P}^{(01,1)1}(0) 
 - 2\sqrt{5}m_{N}\hat{P}^{(11,2)1}(0) = 0 \,.
\end{equation}
Because of the constraints, only six instead of ten GPs are independent.
Though there is some arbitrariness in the selection of the independent
GPs, a natural choice has been proposed in a recent review 
article of Guichon and Vanderhaeghen~\cite{Guichon_98}.
The authors eliminate the three mixed GPs $\hat{P}^{(01,1)0} \,, \; 
\hat{P}^{(01,1)1} \,,$  $\hat{P}^{(11,2)1}$ and the quantity 
$P^{(11,00)1}$. 
In the following discussion we focus on the remaining six GPs.

\subsection{Results and Discussion}
Both nonresonant and resonant excitations of the nucleon contribute to 
the GPs. However,  in the kinematical region of Compton scattering
below pion threshold, the nonresonant s-wave production of pions is of 
particular importance. This process is determined by chiral symmetry and can 
be described
by pion-loop diagrams in ChPT or in the LSM.
On the other side, we also expect large influences of resonances.
In particular, the $\Delta(1232)$ and the $D_{13}(1520)$ resonances,
which are both clearly visible in the photoabsorption spectrum of 
the nucleon, significantly contribute to some of the 
GPs~\cite{Guichon_95,Pasquini_98}.
\\
In addition to these excitations of the nucleon, the chiral anomaly 
($\pi_0$ exchange in
the $t$ channel) is quite essential in the case of the vector GPs.
At low values of $\bar{q}$, the shape for three of the four 
independent vector GPs is strongly governed by the 
anomaly~\cite{Hemmert_97b,Metz_97}, as will be discussed
in sect.~3.

Before comparing the numerical predictions of different approaches,
we discuss the analytical results for the GPs obtained in the
LSM~\cite{Metz_96,Metz_97} and in HBChPT to lowest 
order (${\cal O}(p^3)$)~\cite{Hemmert_97a,Knoechlein_97}.
The chiral expansions of the GPs at $Q_0^2=0$
contain the most instructive information.
As an example, we find for the electric and magnetic polarizabilities of the 
proton
in the case of the one-loop approximation to the LSM
\begin{eqnarray} \label{pol_exp}
\alpha_{p}(0) & = &
 \frac{e^{2}g_{\pi N}^{2}}{192 \pi^{3} m_{N}^{3}}
 \Bigl[ \frac{5\pi}{2\mu} + 18\ln\mu +\frac{33}{2} 
       + {\cal O}(\mu) \Bigr]\nonumber\\
& =& \Bigl[ 13.6-8.8+4.2+{\cal O}(\mu) \Bigr] 
\times 10^{-4}\, \mbox{fm}^3 \,, \nonumber \\
\beta_{p}(0) & = &
 \frac{e^{2}g_{\pi N}^{2}}{192 \pi^{3} m_{N}^{3}}
 \Bigl[ \frac{\pi}{4\mu} + 18\ln\mu +\frac{63}{2} 
       + {\cal O}(\mu) \Bigr]\nonumber\\
& = & \Bigl[1.4-8.8+8.1+{\cal O}(\mu) \Bigr]
\times 10^{-4}\, \mbox{fm}^3\,,
\end{eqnarray}
with $\mu = m_{\pi}/m_{N}$ and 
$g_{\pi N} =m_{N} g_{A}/F_{\pi}\approx 13.4$ 
the pseudoscalar pion--nucleon coupling constant.
It is interesting to note that the chiral expansions of Eq.~(\ref{pol_exp})
agree with a one-loop calculation of 
relativistic ChPT~\cite{Bernard_92} to all orders in $\mu.$ 
The corresponding results of HBChPT to ${\cal O}(p^3)$ 
are given by the leading term in Eq.~(\ref{pol_exp}).
Similar chiral expansions have been obtained for all of the GPs and for their 
derivatives with respect to 
$Q_0^2\,\,$~\cite{Knoechlein_97,Metz_96,Metz_97}. 
Both chiral approaches completely agree in the sense that in all 
cases the leading contribution of the expansion in the LSM is exactly 
the result of HBChPT to ${\cal O}(p^3)$.
\\
The chiral expansions of the polarizabilities show a common 
feature: compared to the leading term the contribution next to leading 
order has a different sign.
Such an alternating behaviour of the chiral series can also be 
observed in HBChPT to ${\cal O}(p^4)$ for $\alpha(0)$ and $\beta(0)$ 
if one neglects counterterm contributions which arise from resonance 
saturation~\cite{Bernard_93}.
\\
The pion cloud of the nucleon causes a diamagnetic
response and therefore a negative magnetic polarizability.
According to Eq.~(\ref{pol_exp}), this 
diamagnetic part of $\beta$ is not contained in the leading chiral
contribution but in the terms next to leading order (logarithmic plus
constant term).  
\begin{figure}
\centerline{
\psfig{figure=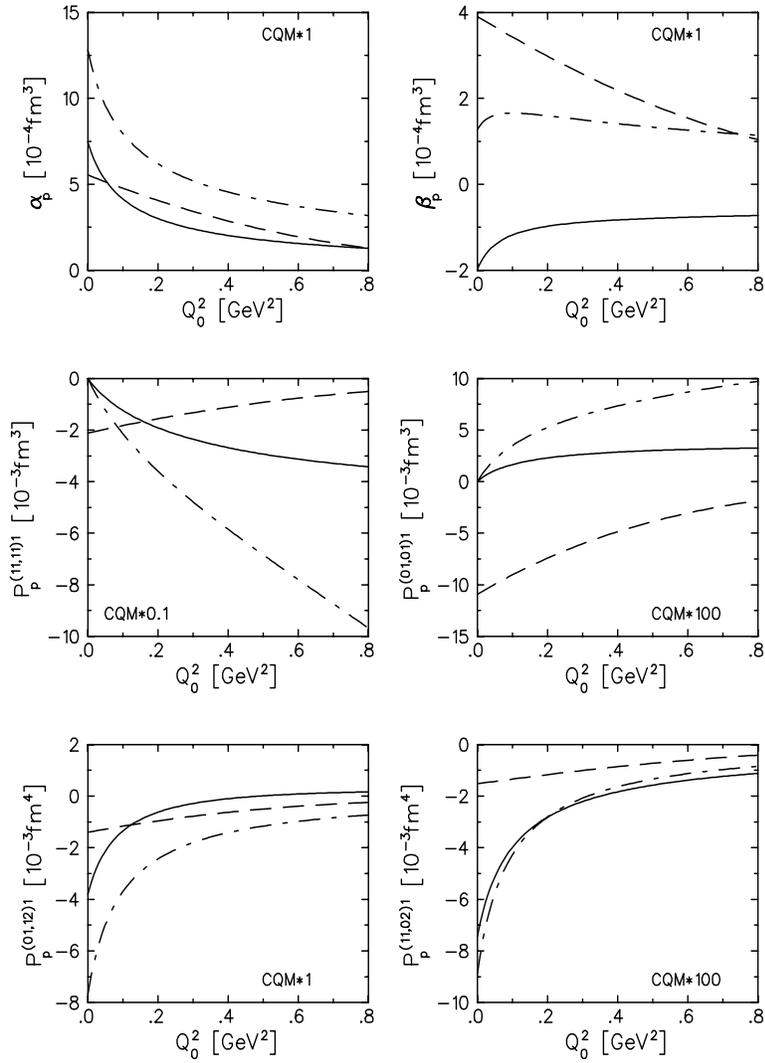,height=14cm,width=10cm}}
\caption{Comparison of GPs as function of $Q_0^2$ obtained in different 
 models. The anomaly is not included.
 Solid line: LSM, dashed line: CQM, 
 dash--dotted line: HBChPT to ${\cal O}(p^3)$.
 Note that the CQM results have been scaled. \label{fig_pol}}
\end{figure}
In Fig.~\ref{fig_pol} we plot the GPs as a function of $Q_0^2$
calculated in the CQM~\cite{Liu_96}, in HBChPT to order 
${\cal O}(p^3)$~\cite{Hemmert_97b,Knoechlein_97} and in the 
LSM~\cite{Metz_96,Metz_97}.
We first consider the scalar polarizabilities $\alpha$ and $\beta$.
At the real photon point ($Q_0^2 = 0$), the LSM underestimates
the empirical values~\cite{Mac_95}
($\alpha_{p}^{exp}(0) = 12.1 \pm 1.0 \times 10^{-4}\,\textrm{fm}^3$,
  $\beta_{p}^{exp}(0) = 2.1 \mp 1.0 \times 10^{-4}\,\textrm{fm}^3$)
to some degree.
This drawback can be attributed to the neglect of t-channel exchange of
 heavier or more meson states, and of nucleon resonances in the s-channel,
e.g. D$_{13}(1520)$ in the case of $\alpha$ and $\Delta(1232)$ with its 
strong paramagnetic (quark spin-flip) contribution to $\beta.$
\\
In HBChPT the results at the real photon point are in good 
agreement with the experiment.
However, at least for the magnetic polarizability, this agreement
is somewhat accidental because both the paramagnetic contribution
of the $\Delta(1232)$ and the main diamagnetic contribution of the pion
cloud have not been taken into account.
The results for $\alpha$ and $\beta$ to ${\cal O}(p^4)$ show a large 
cancellation between diamagnetic and paramagnetic terms~\cite{Bernard_93}, 
where the final results for the proton are 
$\alpha_p=10.5\pm 2.0 \times 10^{-4}\,\textrm{fm}^3$ and 
$\beta_p=3.5\pm 3.6 \times 10^{-4}\,\textrm{fm}^3$. 
In a recent development the $\Delta$ was incorporated in to the 
effective chiral lagrangian.
Unfortunately, this formalism predicts too large 
numbers for $\alpha$ and $\beta$ in a calculation to
${\cal O}(\epsilon^3)$~\cite{Hemmert_97b,Knoechlein_97}.  
\\
Though the values of $\alpha$ and $\beta$ at the real photon
point are different for HBChPT and the LSM, both chiral calculations
predict a very similar $Q_0^2$ behaviour for these polarizabilities. 
In the CQM the $Q_0^2$ dependence is rather different in comparison 
to the chiral calculations, the discrepancy being strongest 
in the case of $\beta$. 
\\
At low $Q_0^2$, the polarizability $P_{p}^{(11,11)1}$ is very large 
in the CQM in contrast to the chiral calculations.
This behaviour is due to the neglect of
diamagnetic contributions in the CQM.
Furthermore,
the CQM predicts finite values for the GPs $P_{p}^{(11,11)1}$ and 
$P_{p}^{(01,01)1}$ at the origin and therefore violates
the model-independent constraints of Eq.~(\ref{const_a}).
The most remarkable discrepancy between the CQM and the chiral 
predictions is in the absolute values of $P_{p}^{(01,01)1}$ and 
$P_{p}^{(11,02)1}$ which differ by two orders of magnitude.   
\section{Observables}
\subsection{Formalism}
In the case of the unpolarized reaction $p(e, e' p' \,) \gamma$ we 
consider the five-fold differential cross section,
\begin{equation}
\frac{d^{5} \sigma}{d|\vec{k}'_{lab}| (d\Omega_{k'})_{lab} 
 (d\Omega_{p'})_{cm}}
 = K_{1} \frac{1}{4} \sum_{spins} |T^{e e' \gamma}|^{2} \,,
\end{equation}
where $K_1$ represents a phase space factor.
The cross section depends on the five variables
\begin{equation}
\omega'\,, \quad \bar{q} \,, \quad \theta \,, \quad
\phi \,, \quad \epsilon \,,
\end{equation}
with $\theta$ the angle between the two photons, 
$\phi$  the azimuthal angle between the leptonic plane and the reaction
plane, and $\epsilon$ the transverse polarization of the virtual photon.
According to Eqs.~(\ref{interf}) and (\ref{decomp}), the scattering amplitude
$T^{e e' \gamma}$ consists of three parts,
\begin{equation}\label{totamp}
T^{e e' \gamma} = T^{BH} + T_{B}^{VCS} + T_{R}^{VCS} \,.
\end{equation}
As explained in sect.~1, the "background" amplitudes
$T^{BH}$ and $T_{B}^{VCS}$ behave like $1/\omega'$ in the
low energy limit, while $T_{R}^{VCS}$ is proportional to $\omega'$.
Therefore, the expansion of the spin-averaged matrix element reads
\begin{equation} \label{gl_ampexp}
\frac{1}{4} \sum_{spins} |T^{e e' \gamma}|^{2} =  
 \frac{C^{BH+Born}_{-2}}{\omega^{\prime 2}} + \frac{C^{BH+Born}_{-1}}{\omega'}
 + C^{BH+Born}_{0} + C^{Pol}_{0}+ {\cal O}(\omega')\,,
\end{equation}
where the coefficients $C^{BH+Born}_{-2},$ 
$C^{BH+Born}_{-1}$ and $C^{BH+Born}_{0}$ are entirely determined
by the background. 
The coefficient $C^{Pol}_{0}$ results from the interference of the 
singular part of the background term with the leading term of 
$T_{R}^{VCS}$ and contains the information on the GPs.
It can be 
expressed in terms of 
four structure functions~\cite{Guichon_95},
\begin{eqnarray} \label{int_unpol}
C_{0}^{Pol} & = & 
 K_{2}(\epsilon, \bar{q}) \Bigl[
 v_{1} \bigl(\epsilon P_{LL}(\bar{q}) - P_{TT}(\bar{q}) \bigr)
+ v_{2} \sqrt{2\epsilon(1+\epsilon)} P_{LT}(\bar{q})
 \nonumber \\
& & \hspace{1.5cm}
+ v_{3} \sqrt{2\epsilon(1+\epsilon)} P_{LT}'(\bar{q}) \Bigr] \,,
\end{eqnarray}
with kinematical factors $v_{i}$ depending on $\bar{q}$,
$\theta$, $\phi$ and $\epsilon$.
The structure functions contain the GPs in combination with the 
elastic form factors of the nucleon. 
Using the results of Ref.~\cite{Drechsel_98a}, the structure
functions $P_{LT}$ and $P'_{LT}$ are mutually dependent via the relation
\begin{equation}
P_{LT}(\bar{q}) = \frac{2 m_{N} \bar{q}}{Q_{0}^{2}} P_{LT}'(\bar{q}) \,.
\end{equation}
As a consequence, the coefficient $C^{Pol}_0$ contains only three 
independent structure functions, thus leading to information on 
three of the six independent GPs.

In order to disentangle the remaining polarizabilities one has to 
resort to double polarization observables.
As has been suggested by Vanderhaeghen~\cite{Vanderhaeghen_97},
the reaction $p(\vec{e}, e' \vec{p}\,')\gamma$ can be used to extract the
remaining three polarizabilities.
Instead of measuring recoil polarization, the same information 
can be gained by the reaction $\vec{p}(\vec{e}, e' p'\,) \gamma$.
We focus the attention on the observables for beam-target polarization
in order to give some guidance for a possible experiment with the BLAST 
facility at MIT-Bates. 
To this end we consider the target (double) asymmetries  
\begin{equation} \label{eq_asy}
T^{(i)} = \frac{ [ \sigma_{h=\frac{1}{2}, m_{i}=\frac{1}{2}}
   - \sigma_{h=\frac{1}{2}, m_{i}=-\frac{1}{2}}]
 - [\sigma_{h=-\frac{1}{2}, m_{i}=\frac{1}{2}}
   - \sigma_{h=-\frac{1}{2}, m_{i}=-\frac{1}{2}}]}
 { [\sigma_{h=\frac{1}{2}, m_{i}=\frac{1}{2}}
   + \sigma_{h=\frac{1}{2}, m_{i}=-\frac{1}{2}}]
   + [\sigma_{h=-\frac{1}{2}, m_{i}=\frac{1}{2}}
   + \sigma_{h=-\frac{1}{2}, m_{i}=-\frac{1}{2}}]} \,,
\end{equation}
where $h$ is the electron helicity and $m_{i}\,\, (i=x,y,z)$ is the 
spin-projection of the nucleon target with respect to the coordinate 
system
\begin{equation}
\hat{e}_{x} = \frac{\hat{q}' - \cos \vartheta \, \hat{q}}
 {\sin\vartheta}\,, \quad
\hat{e}_{y} = \frac{\hat{q} \times \hat{q}'}{\sin \vartheta} \,, \quad
\hat{e}_{z} = \hat{q} \,.
\end{equation}
As in the unpolarized case, 
one 
can perform a low energy expansion for the numerator of the asymmetries 
leading to expansion coefficients
$C_{0(i,0)}^{Pol}$.
In terms of structure functions these coefficients read 
\begin{eqnarray} \label{int_polz}
C_{0(z,0)}^{Pol} & = & 4 K_{2}(\epsilon,\bar{q}) \Big[
 v_{1} \sqrt{1-\epsilon^{2}} P_{TT}^{(z,0)}(\bar{q})
+ v_{2} \sqrt{2\epsilon (1-\epsilon)} P_{LT}^{(z,0)}(\bar{q}) 
\nonumber \\
& & \hspace{1.5cm} + v_{3} \sqrt{2\epsilon (1-\epsilon)} 
 P_{LT}^{\prime(z,0)}(\bar{q}) \Big] \,,
 \\ \label{int_polx}
C_{0(x,0)}^{Pol} & = & 4 K_{2}(\epsilon,\bar{q}) \Big[
 v_{1}^{x} \sqrt{2\epsilon (1-\epsilon)} P_{LT}^{(\perp,0)}(\bar{q})
+ v_{2}^{x} \sqrt{1-\epsilon^{2}} P_{TT}^{(\perp,0)}(\bar{q}) 
\nonumber \\
& & \hspace{1.5cm}+ v_{3}^{x} \sqrt{1-\epsilon^{2}} 
 P_{TT}^{\prime (\perp,0)} (\bar{q}) 
 + v_{4}^{x} \sqrt{2\epsilon (1-\epsilon)} P_{LT}^{\prime (\perp,0)}
 \Big] \,. 
\end{eqnarray} 
A similar decomposition of $C_{0(y,0)}^{Pol}$ contains the same structure 
functions as $C_{0(x,0)}^{Pol}$ in Eq.~(\ref{int_polz}), however with
new kinematical factors $v_{i}^{y}$.
Note that the kinematical factors for target polarization in $z$-direction
are the same as in the unpolarized case.
Altogether one obtains three new independent structure functions,
$P_{LT}^{(z,0)}$, $P_{LT}^{\prime(z,0)}$ and $P_{LT}^{\prime(\perp,0)}$.
The extraction of $P_{LT}^{\prime(\perp,0)}$ requires an out-of-plane 
measurement since $v_4^x \to 0$ for $\phi \to 0^{{\rm o}}$.
This situation is quite analogous to the case of double polarization 
asymmetries with recoil polarization~\cite{Vanderhaeghen_97},
whose expansion coefficients contain the same kinematical factors as 
for target polarization,
while the corresponding structure functions may have a 
somewhat different appearance.
The explicit expressions for the coefficients $C_{0(i,0)}^{Pol}$ will be 
presented in an upcoming publication~\cite{Drechsel_98c}.

\subsection{Results and Discussion} 
In the following numerical studies we concentrate on two 
different kinematics: (I) the Bates kinematics with 
$\bar{q} = 240 \,\textrm{MeV}$, $\epsilon = 0.9$ and 
$\omega' = 100 \,\textrm{MeV}$ and (II)
the MAMI kinematics with
$\bar{q} = 600 \,\textrm{MeV}$, $\epsilon = 0.62$ and
$\omega' = 111 \,\textrm{MeV}$.
Both sets of variables have been explored in the unpolarized 
experiments at Bates~\cite{Bates_prop} and MAMI~\cite{MAMI_prop}.
The main difference beteween the two kinematics is in the values of $\bar{q}$ 
and $\epsilon$, while the difference in $\omega'$ will be neglected in 
the discussion below. 
\\
In our analysis, the observables are always calculated as function 
of the scattering angle $\theta$. 
In addition,  we distinguish between in-plane
($\phi=0$) and the $90^{{\rm o}}$ degree out-of-plane kinematics
for the Bates kinematics,
while in the case of the MAMI kinematics the in-plane and $60^{{\rm o}}$ 
degree out-of-plane cases have been investigated.
\begin{figure}
\centerline{
\psfig{figure=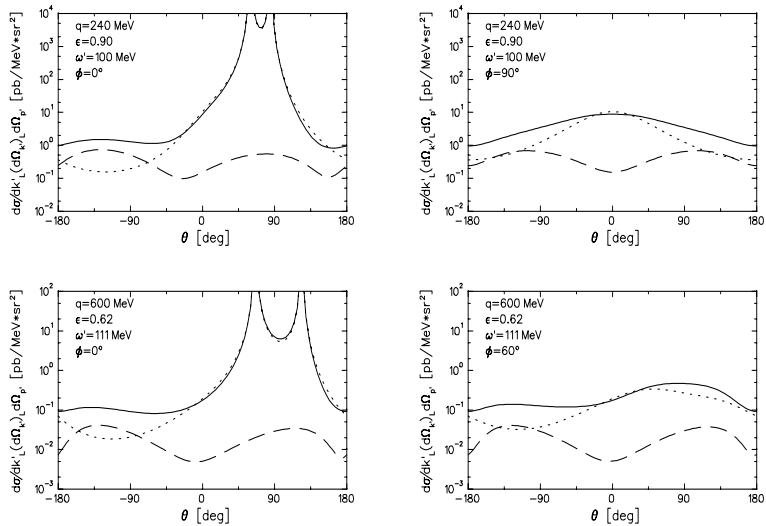,width=10cm,height=6.9cm,angle=90}}
\caption{Background cross section for the Bates kinematics and 
the MAMI kinematics as function of the scattering angle $\theta.$
Dashed line: VCS Born contribution, dotted line: Bethe-Heitler
contribution, full line: complete background cross section.
 \label{fig_back}}
\end{figure}
Results for the background cross section are shown in
Fig.~\ref{fig_back}, with the separate contributions from
the BH and the VCS Born amplitude.
For the elastic form factors of the proton we used the parametrization 
of Ref.~\cite{Mergell_96}. 
In the case of in-plane kinematics, the angular dependence of the 
BH cross section is characterized by two peaks occurring for the real 
photons emitted along the direction of the initial or final electrons.
In these regions, the BH term gives the overwhelming contribution to
the background cross section, while it becomes much smaller at 
negative values of $\theta$. 
For out-of-plane configurations the two peaks disappear, and in the 
case of the Bates kinematics the angular distribution of the background term 
is almost symmetrical around $\theta=0$.
In order to increase the sensitivity to the GPs, the region of negative 
scattering angles corresponding to photon emission in the half-plane 
opposite to the outgoing electron has to be explored.
In comparison to the MAMI kinematics, the background cross section is one 
order of magnitude higher in the Bates case, whereas the $\theta$
dependence of the background is similar for both kinematics. 

\begin{figure}[t]
\centerline{
\psfig{figure=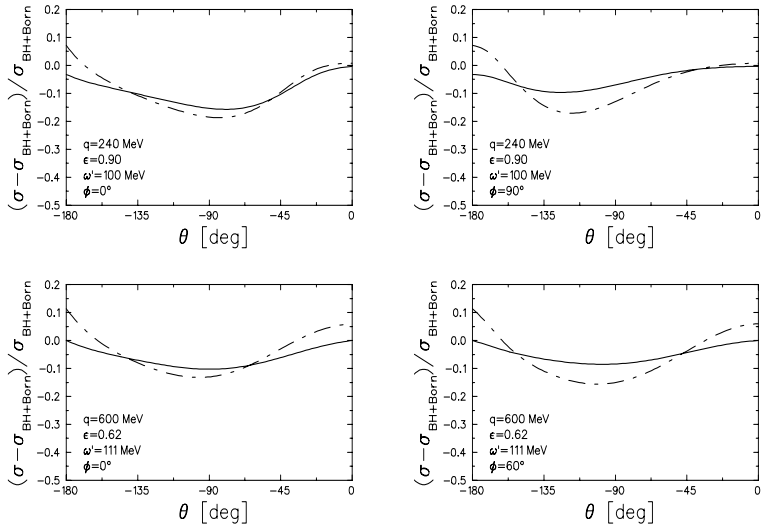,width=10cm,height=6.9cm,angle=90}}
\caption{Influence of the GPs on the unpolarized cross section
 of the reaction $p(e, e' p' \,) \gamma$ for Bates and MAMI kinematics.
 Full line: LSM result, dash-dotted line: HBChPT result.
 The quantity $\sigma - \sigma_{BH+Born}$ represents the contribution
 of the coefficient $C_{0}^{Pol}$ of Eq.~(\ref{int_unpol}) to the cross
 section.
 \label{fig_unpol}}
\end{figure}

\begin{figure}[!tbp]
\centerline{
\psfig{figure=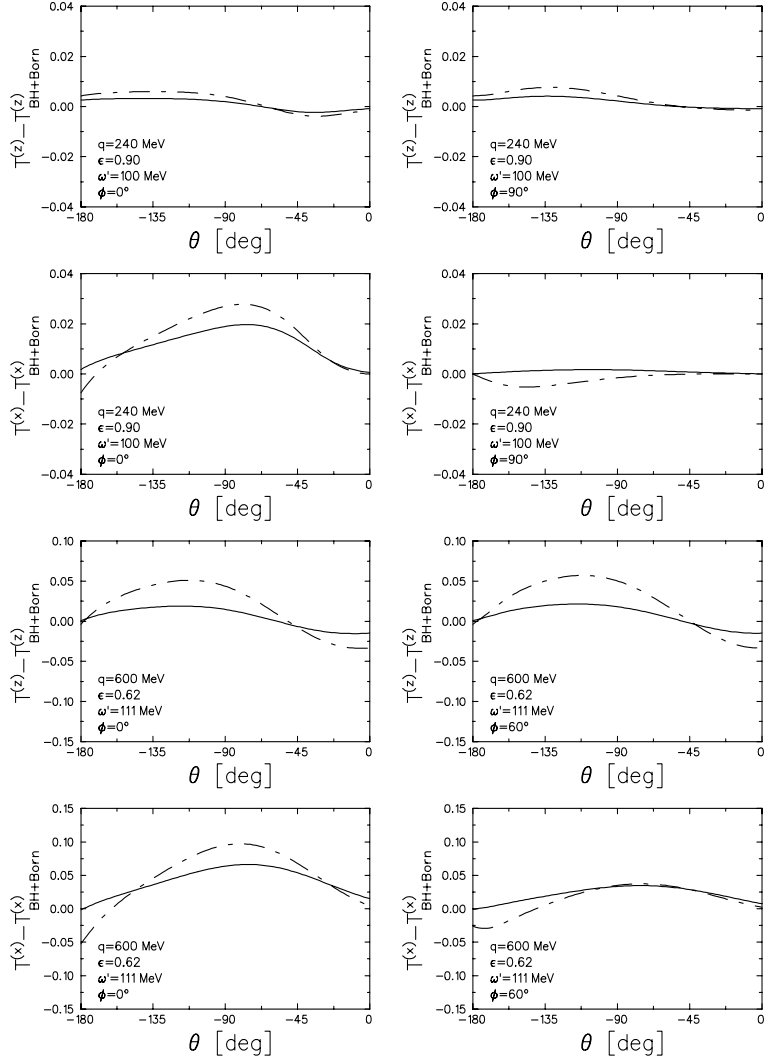,height=14cm,width=10cm}}
\caption{Influence of the GPs on the target asymmetries for Bates 
 and MAMI kinematics.
 Full line: LSM result, dash-dotted line: HBChPT result. 
 The quantities $T^{(z)} - T_{BH+Born}^{(z)}$ and
 $T^{(x)} - T_{BH+Born}^{(x)}$ represent the contribution of the
 coefficients $C_{0(z,0)}^{Pol}$ and $C_{0(x,0)}^{Pol}$ in 
 Eqs.~(\ref{int_polz}) and (\ref{int_polx}), respectively.
 \label{fig_asy}}
\end{figure}

For the LSM and HBChPT the relative influence 
of the GPs on the unpolarized cross section has been
plotted in Fig.~\ref{fig_unpol}.  
As is to be expected, HBChPT generally predicts larger effects than the LSM. 
In both calculations the influence of the GPs is negative for a large 
region of $\theta$, leading to a reduction of the total cross section.
For the unpolarized case the anomaly gives no contribution 
as long as only the leading order in $\omega'$ is taken into account.
\\
For the Bates kinematics the GPs change the cross section by about 20\%, 
the effect being almost completely given by $\alpha(\bar{q})$.
The influence of the GPs for the MAMI kinematics is of the order of 15\%. 
In this case the LSM predicts that  only 60\% of the signal are due to the 
electric polarizability.
Since one aims at cross section measurements with a relative error
of about 2\%, the calculated effects should be detected.
However, there are two problems which render the measurement of
the GPs more difficult: (I)
the effect of the GPs strongly increases with $\omega'$.
At $\omega' = 70 \,\textrm{MeV},$ e.g., the effects shown in 
Fig.~\ref{fig_unpol} reduce by roughly 50\%.
For higher values of $\omega'$, where the signal is larger, higher 
order terms of the residual amplitude can no longer be neglected
(see also the corresponding discussion in sect.~2.1).
In this case the challenging task is the reliable extraction of the 
leading order contribution in which one is interested. 
(II) Independent of the value of $\omega'$ there are large radiative 
corrections~\cite{Vanderhaeghen_98}. 
These corrections result in a 20\% contribution to the cross section
which is of the same order of magnitude as the influence of the GPs,
thus complicating the separation of background and GP effects.
\\
In Fig.~\ref{fig_asy} we show the GP contribution to the target 
asymmetries $T^{(z)}$ and $T^{(x)}$ for both LSM and HBChPT, 
without taking account of the anomaly contribution.
Obviously, larger effects can be expected in the case of $T^{(x)}$.
While in the unpolarized cross section the GP effect was larger 
for Bates kinematics, we now observe the opposite situation for the
asymmetries.
Here the maximum signature is about 10\% for the MAMI kinematics but
only about 3\% for the Bates kinematics.
The different behaviour between the unpolarized and the polarized case 
has the following explanation:
In Fig.~\ref{fig_unpol} we have seen that the background, which determines 
the denominator of the asymmetries, is one order of magnitude larger
for the Bates kinematics.
On the other side, the influence of the GPs in the numerator of the
asymmetries increases much less when going from Bates to MAMI 
kinematics, while the influence of the GPs on the unpolarized 
cross section also strongly increases by changing the kinematics,
in particular due to the strong $\bar{q}$ dependence of the electric 
polarizability.
More details of this discussion will be presented in~Ref.~\cite{Drechsel_98c}.
\begin{figure}[!tbp]
\centerline{
\psfig{figure=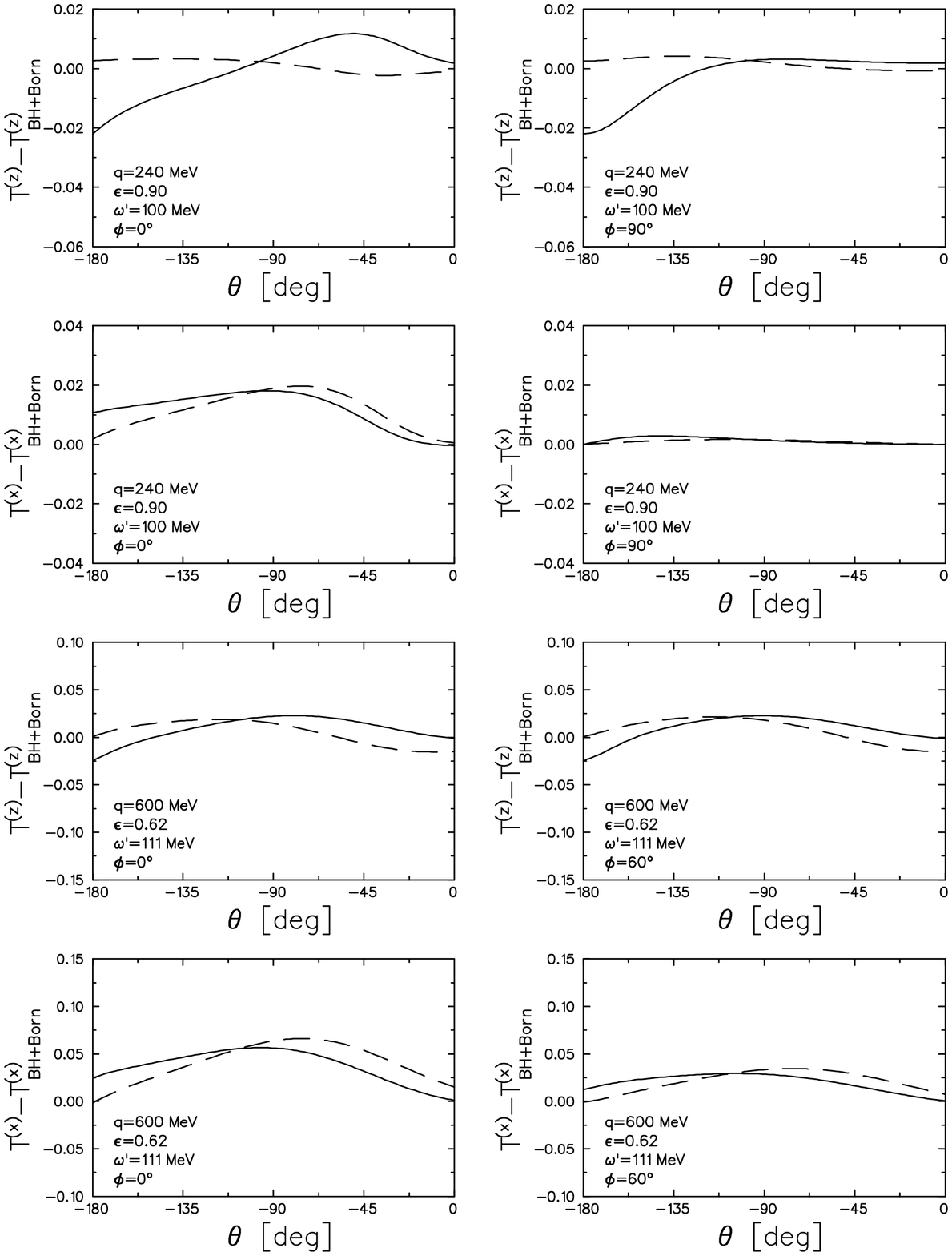,height=14cm,width=10cm}}
\caption{Target asymmetries with and without the anomaly for Bates and 
 MAMI kinematics as obtained in the LSM.
 Full line: anomaly included, dashed line: without anomaly.
 \label{fig_asy_anom}}
\end{figure}
\\
In Fig.~\ref{fig_asy_anom} we display the anomaly contribution to the 
asymmetries as predicted by the LSM.
Since the influence of the anomaly is strongly decreasing with
increasing $\bar{q}$, the largest effect shows up for the Bates 
kinematics.
In this case the shape of the asymmetry $T^{(z)}$ is completely changed
by the anomaly.
Though for the MAMI kinematics the anomaly is less dominant,
its influence is still quite sizeable.

We conclude that a full determination of the GPs will require double 
polarization experiments, polarized electrons and target or recoil 
polarization, in addition to the unpolarized experiments.
Given the projected accuracy of the experiments, one should be able
to measure the GPs in spite of a large background. Since the momentum 
dependence of the GPs is predicted quite differently by various models,
such precision experiments will be invaluable to restrict the model 
parameters, and, quite generally, as further benchmarks of the 
nucleon's structure.
\section*{Acknowledgments}
The authors thank M.~Vanderhaeghen for useful discussions concerning 
the formalism of the observables.

\end{document}